**Resource Letter RBAI-2:**
**Research-based assessment instruments: Beyond physics topics**

Adrian Madsen, Sarah B. McKagan

*American Association of Physics Teachers, College Park, MD*

Eleanor C Sayre

*Department of Physics, Kansas State University, Manhattan, KS 66506*

Cassandra A. Paul

*San Jose State University, San Jose, CA*

**Abstract**

This Resource Letter provides a guide to research-based assessment instruments (RBAIs) for physics and astronomy classes, but goes beyond physics and astronomy topics to include: attitudes and beliefs about physics, epistemologies and expectations, the nature of physics, problem solving, self-efficacy, reasoning skills and lab skills. We also discuss RBAIs in physics and astronomy cognate fields such as mathematics and observation protocols for standardized observation of teaching. In this Resource Letter, we present an overview of these assessments and surveys including research validation, instructional level, format, and themes, to help faculty find the assessment that most closely matches their goals. This Resource Letter is a companion to RBAI-1: Research-based Assessment Instruments in Physics and Astronomy, which explicitly dealt with physics and astronomy topics. More details about each RBAI discussed in this paper are available at PhysPort: physport.org/assessments.

**I. INTRODUCTION**

In the first Resource Letter in this series (RBAI-1),[1] we presented 40+ Research-

Based Assessment Instruments (RBAIs) developed and used by the physics and astronomy education community to assess student understanding of physics and astronomy content. Here, we present RBAIs used by the physics, astronomy and science education research communities to examine non-physics/astronomy-content related topics. In our interviews with physics faculty, we have learned that faculty want to assess not only students' conceptual learning, but the skills and attitudes that faculty hope students gain as a result of their physics/astronomy course.[2] This Resource Letter is meant to help faculty find and use the assessments that are applicable to their students and their goals beyond content.

As we look at RBAIs beyond physics and astronomy topics, we had to decide what to include and not to include because the space of cognate fields is quite large, and researchers in discipline-based education research and the psychological sciences have developed hundreds of RBAIs over decades. In the Resource Letter, we focused on the RBAIs that physics and astronomy faculty are likely to find most helpful in the course of teaching and assessing their classes. Towards that end, we have not included instruments for programmatic assessment or generalized student development, as well as instruments intended primarily for use by researchers.

We begin with a general discussion of the RBAIs included in this Resource Letter and their research validation (Section II), and then discuss specific RBAIs in several major categories. These RBAIs cover a diverse set of topics including mathematics (Section III), attitudes and beliefs (including nature of science and self-efficacy) (Section IV), problem-solving (Section V), scientific reasoning (Section VI), lab skills (Section

VII), and observation protocols (Section VIII), at a range of levels from high school to graduate school.

More details about each of these RBAIs are available at physport.org/assessments,[3] where verified educators can download most of the RBAIs.

1. "Resource Letter RBAI-1: Research-Based Assessment Instruments in Physics and Astronomy," A. Madsen, S. B. McKagan, and E. C. Sayre, Am. J. Phys. **85**(4), 245–264 (2017). (E)

2. "Research-based assessment affordances and constraints: Perceptions of physics faculty," A. Madsen, S. B. McKagan, M. "Sandy" Martinuk, A. Bell, and E. C. Sayre, Phys. Rev. - Phys. Educ. Res. **12**(010115), 1–16 (2016). (E)

3. "PhysPort Assessments," www.physport.org/assessments. (E)

## II. RESEARCH-BASED ASSESSMENTS INSTRUMENTS BEYOND PHYSICS TOPICS

Good research-based assessment instruments are different from typical exams in that their creation involves extensive research and development by experts in education research to ensure that the questions measure the constructs that faculty think are important, that the possible responses represent real student thinking and make sense to students, and that students' scores reliably tell us something about their understanding. For an overview of the development process for research-based assessments, see Madsen et al., 2017.[1]

The assessments and observation protocols discussed in this Resource Letter are all developed using research-based approaches, but because most of them are not assessments of physics or astronomy topics, there are differences in how they are developed, structured, and used.

In the first Resource Letter in this sequence, RBAI-1, most of the assessments were designed to be offered before and after instruction, allowing faculty to assess their instruction by comparing the gain between pre- and post-test scores. For the RBAIs in this Resource Letter, there is a much larger variety in the format, administration, and interpretation of results. Measuring students' skills or beliefs is more difficult than measuring their conceptual knowledge. Developers don't just build multiple-choice questions. Instead they use a variety of assessment formats including asking students to agree or disagree with statements, rubrics to assess a certain skill, open-ended responses, choosing multiple responses, or not scoring questions at all, but instead just discussing answers.

To score beliefs assessments, students are often compared to a normative group of physics experts. For example, if experts disagree with a statement that physics is about memorizing information, then students who also disagree may earn one point, while students who agree with that statement do not. The overall score is a measure of how much students agree with physicists. To track changes over time, we look at "shifts" in students' scores.

Rubrics and observation protocols are usually scored by identifying response patterns or behaviors that are present (or absent) and assigning points to their presence (or absence).   The amount of points can vary by which rubric or observation protocol you are using and what it focuses on.  This is a substantially different scoring system than asking students to fill in a bubble sheet: the person who is scoring the student work or observing the class makes judgments about the work or behavior they observe.  Rubrics help faculty and students understand the students' strengths and areas for growth for a variety of categories, while observation protocols help faculty understand the activities in classrooms for a variety of settings and activities. They can both be powerful forms of formative assessment for both the students and faculty.

**TABLE I.** Research validation categories for content and beliefs RBAIs as well as observation protocols.

| Categories for content, beliefs, and reasoning RBAIs | Categories for observation protocols |
|---|---|
| Questions based on research into student thinking | Categories based on research into classroom behavior |
| Studied with student interviews | Studied using iterative observations |
| Studied with expert review | Tested using inter-rater reliability |
| Appropriate use of statistical analysis | Training materials are tested |
| Administered at multiple institutions | Used at multiple-institutions |
| Research published by someone other than developers | Research published by someone other than developers |
| At least one peer-reviewed publication | At least one peer-reviewed publication |

We determine the level of research validation for an assessment based on how many of the research validation categories apply to the RBAI (Table I and Table II).  RBAIs will have a gold level validation when they have been rigorously developed and recognized by a wider research community. Silver-level RBAIs are well-validated, but are missing some piece, such as validation by the larger community. Bronze-level assessments are those where developers have done some validation but are missing several pieces. Finally, a bronze research-based validation means that an assessment is likely still in the early stages. We have developed separate levels of research validation for observation protocols because the development process for these is substantially different than for the other kinds of assessments. Because faculty, and not students, use protocols, it does not make sense to look at student thinking or do student interviews. Instead, when developing observation protocols, it is vital to ensure that the categories of observation are grounded in real classrooms. The protocol is iteratively developed through use in real classrooms, there is a high level of inter-rater reliability (which means the observers can interpret and apply the protocol similarly) and the training materials for using the protocol have been tested and refined. To reflect the differences between observation protocols and other types of RBAIs, we developed a parallel set of research validation categories for observation protocols (Table I).

**TABLE II**. Determination of the level of research validation for an assessment.

| # Categories | Research validation level |
|---|---|
| All 7 | Gold |
| 5-6 | Silver |
| 3-4 | Bronze |
| 1-2 | Research-based |

## III. MATHEMATICS ASSESSMENTS

**TABLE III**. Mathematics assessments.

| Assessment | Content | Intended population | Research validation | Purpose |
|---|---|---|---|---|

| Calculus Concept Inventory (CCI) | Derivatives, functions, limits, ratios, the continuum | Intro college, high school | Gold | Assess student understanding of the most basic principles of calculus. |
|---|---|---|---|---|
| Pre-calculus Concept Assessment (PCA) | Rate of change, function, process view of functions, covariational reasoning | Intro college | Gold | Assess essential knowledge that mathematics education research has revealed to be foundational for students' learning and understanding of the central ideas of beginning calculus. |
| Calculus Concept Readiness (CCR) | The function concept, trigonometric functions and exponential functions | Intro college | Silver | Assess the effectiveness of pre-calculus level instruction or to be used as a placement test for entry into calculus. |
| Test of Understanding of Vectors (TUV) | Vectors, components, unit vector, vector addition, subtraction and multiplication, dot and cross product | Intro college | Silver | Assess students' understanding of vector concepts in problems without a physical context. |
| Quadratic and Linear Conceptual Evaluation (QLCE) | Graphing, mathematical modeling | Intro college, high school | Bronze | Measure student understanding of equations (linear and quadratic) as functional relationships. Also, to measure students' mathematical knowledge in both traditional and reform courses. |
| Vector Evaluation Test (VET) | Vector addition and subtraction, component analysis, and comparing magnitudes | Intro colleges high school | Bronze | Measure students' conceptual understanding of vectors. |

## A. Overview of Mathematics Assessments

RBAIs for mathematics can be used in physics classes to assess students' level of math readiness for a given physics class, or to assess students' understanding of math topics that are covered in physics classes. These tests are often used in concert with or instead of mathematics placement exams developed locally. We discuss three mathematics assessments, developed by mathematics education researchers, that you can use before instruction to get a sense of students' prerequisite mathematics skills and to assess calculus readiness. You could also use these as pre/post-tests to see how your students' calculus skills improved because of your course. These are the Precalculus Concept Assessment[4] (PCA), the Calculus Concept Inventory[5,6] (CCI), and the Calculus Concept Readiness Instrument[7] (CCR). There are three additional assessments, developed by physicists, that assess math topics often taught in physics classes, i.e., vectors and mathematical modeling. These are the Quadratic and Linear Conceptual Evaluation[8] (QLCE), the Test of Vectors[9] (TUV) and the Vector Evaluation Test[8] (VET). You can use these as a pre- and post-test, to both get a sense of what your students know at the start of your course, and what they learned because of your course. Other tests exist (e.g., the Basic Skills Diagnostic Test[10] (BSDT)), but there is no published information available about them, their developers are unavailable for consultation, and/or we cannot access the assessments.

The Precalculus Concept Assessment[4] (PCA) is a multiple-choice pre/post assessment of foundational concepts of beginning calculus, including reasoning abilities around the process view of functions, covariational reasoning and computational abilities, understanding of the meaning of function concepts, growth rate of function types, and function representations. The PCA can be used to help a physics faculty member understand their student's calculus readiness. The PCA questions were developed based on a taxonomy of precalculus concepts (The PCA Taxonomy) using an iterative process of developing questions, testing them with students, interviewing students about their responses and revising the questions and answer choices.

The Calculus Concept Inventory[5,6] (CCI) is a multiple-choice pre/post assessment of the most basic principles of calculus, where questions are conceptual with no computation on the test. The topics covered on the CCI include functions, derivatives, limits, ratios, and the continuum. The CCI was modeled closely after the Force Concept Inventory[11] (FCI), where the questions look trivial to experts, but students in lecture courses score quite poorly on the test. The CCI questions were first developed by a panel of experts who defined the content to be tested and wrote the questions, and then tested iteratively with students and revised. The CCI is not available for download from PhysPort, because we have not been able to access it ourselves.

While the PCA was developed using a research-based taxonomy of concepts, the CCI was designed to mimic the FCI. This difference means that students' responses to CCI questions are more likely to cause cognitive conflict in physics faculty ("they should have gotten that!") while PCA questions are more likely to present a robust and varied sense of students' understanding of function concepts in a classroom. The CCI is designed for more advanced math skills than the PCA and may be inappropriate for students enrolled in conceptual or algebra-based physics classes; however, in courses which require substantial calculus or differential equations (e.g., intermediate mechanics) it may be a more appropriate pre-test.

The Calculus Concept Readiness[7] (CCR) instrument is a multiple-choice pre/post assessment of foundational concepts for introductory calculus, including the function concept, trigonometric functions and exponential functions. The CCR was developed to assess students' readiness for calculus courses, or to assess the effectiveness of pre-calculus courses. Like the PCA, the CCR was developed using a research-based taxonomy of concepts. The CCR is owned by the Mathematical Association of America and is available for a fee through Maplesoft.[12]

At first blush, the PCA, CCR, and CCI cover very similar topics at a very similar level. However, their emphasis is different, and care should be taken to match the test with your students. The CCR surveys students' understanding of a broad base of mathematics concepts from pre-calculus, including both functions and trigonometry, while the PCA focuses only on the mathematics needed to move into calculus (primarily functions) before calculus instruction. The CCI is designed to test the core concepts of calculus and is aimed at students before and after calculus instruction. Both the PCA and the CCR were developed by the same team of researchers using very similar development methods, and the tests have very similar structure and feel. The CCI was independently developed by a different team using less robust research methods. If you use these as part of a mathematics placement package or to measure their students' mathematics skills, the CCR is recommended because of the trigonometry and solving equations cluster, though you must pay to use it. Physicists are typically not as interested as mathematicians in the intricacies of student understanding of functions, so the PCA and CCI may not be as helpful as the CCR for these purposes.

The Quadratic and Linear Conceptual Evaluation[8] (QLCE) is a multiple-choice assessment about relating kinematics to quadratic graphs and equations, relating coefficient changes in linear equations to linear graph changes and vice versa. Some questions have a kinematics context, and some questions have a generic context. The developers created the QLCE because they had heard faculty say that their students "understood the math, but not the concepts," and wanted to see if their physics students did indeed understand these mathematical concepts. There are several sets of questions where students fill in a matrix to answer, so you would need to renumber them for use with Scantron and will need a special Scantron sheet that can take up to 10 answers and multiple responses for each question. These questions were developed based on research into student ideas about quadratic and linear equations, and the developers' experience with the concepts with their students.

The Test of Vectors[9] (TUV) is a multiple-choice test that assesses introductory physics students' understanding of vector concepts without any physical context. Concepts tested include unit-vector notation, graphical representation of vectors and components, calculation of vector components, vector addition, subtraction and scalar multiplication, and dot and cross product. The TUV questions were developed from students' open-ended responses to questions about vectors, so the multiple-choice answers strongly reflect students' ideas about vectors (both correct and incorrect). The TUV was developed in Mexico in Spanish, and then translated into English.

The Vector Evaluation Test[8] (VET) is a multiple-choice, multiple-response (can pick more than one option) and open-ended assessment of vector concepts for introductory physics classes. About a quarter of the questions are asked in a physics context, and the rest are given no physical context. The VET questions were based on the developers' experience with students thinking about vectors.

Both the TUV and VET cover vector decomposition, addition, subtraction, dot products, and cross products, which are the major issues for using vectors in introductory physics. Additionally, the TUV uses both graphical representations and vector-hat representations, so it is possible to compare students' performance across representations. The VET covers coordinate rotation and time changes of kinematics vectors, so it is more appropriate to use this test if you would like to test more topics instead of more representations. Though it is a more thorough test of the topics it does cover, the TUV's reliance on few questions per topic means that scores are still sensitive to the peculiarities of the questions on the test.

## B. Recommendations for Choosing a Mathematics Assessment

You can use these math assessments before instruction to get a sense of what your students already know, or after instruction if you are implementing new teaching practices to increase students understanding of a given topic and want to assess the effectiveness. Because the QLCE, PCA, CCR, and CCI test overlapping concepts, you should select one of these four that best matches your population and assessment needs. Don't mix-and-match these tests for pre- and post-test because you will have difficulty comparing pre-scores to post-scores.  If you are using a test only before instruction to see if your students are ready to take your course or to adjust your teaching to best fit their incoming skills, select a test of more elementary content that might be fully covered in pre-requisite classes.  If you are using a test before and after instruction, you might select a test that includes some content covered in co-requisite courses.

## IV. BELIEFS AND ATTITUDES

**TABLE IV.** Beliefs and attitudes assessments.

| Title | Focus | Intended population | Research validation | Purpose |
|---|---|---|---|---|
| **Beliefs About Physics Learning in General** | | | | |
| Colorado Learning Attitudes about Science Survey (CLASS) | Self-reported beliefs about physics and learning physics | Upper-level, intermediate, intro college, high school | Gold | Measure students' beliefs about physics and learning physics and distinguish the beliefs of experts from those of novices. |
| Maryland Physics Expectations Survey (MPEX) | Beliefs about one's physics course | Upper-level, intermediate, intro college, high school | Gold | Probe some aspects of student expectations in physics courses and measure the distribution of student views at the beginning and end of the course. |
| Epistemological Beliefs Assessment for Physical Sciences (EBAPS) | Epistemological beliefs, structure of knowledge, nature of knowing and learning, real-life applicability, evolving knowledge, source of ability to learn | Intro college, high school | Silver | Probe the epistemological stances of students in introductory physics, chemistry and physical science. |
| Views About Science Survey (VASS) | Structure and validity of scientific knowledge, scientific methodology, | Intro college, high school | Silver | Characterize student views about knowing and learning science and assess the relation of these views to achievement in science courses. |

learnability of
science, reflective
thinking, personal
relevance of science

**Beliefs About Physics Learning in a Specific Context**

| | | | | |
|---|---|---|---|---|
| Colorado Learning Attitudes about Science Survey for Experimental Physics (E-CLASS) | Affect, confidence, math-physics-data connection, physics community, uncertainty, troubleshooting, argumentation, experimental design, modeling | Upper-level, intermediate, intro college | Gold | Measure students' epistemologies and expectations around experimental physics. |
| Attitudes and Approaches to Problem Solving Survey (AAPS) | Attitudes about problem-solving | Graduate, upper-level, intermediate, intro college | Silver | Measure students' attitudes and approaches to problem solving at the introductory and graduate level. |
| Physics Goals Orientation Survey (PGOS) | Goal orientation and motivation in physics | Intro college | Silver | Assess students' motivation and goal orientations in university-level physics courses. |
| Student Assessment of Learning Gains (SALG) | Self-assessment of learning | Intro college | Silver | Understand students' self-assessment of their learning from different aspects of the course and their gains in skills, attitudes, understanding of concepts, and integrating information. |
| Attitudes about Problem Solving Survey (APSS) | Attitudes about problem-solving | Intro college | Bronze | Survey students' attitudes towards and views of problem solving. |

**Nature of Science**

| | | | | |
|---|---|---|---|---|
| Views of the Nature of Science (VNOS) | Nature of science, theories and laws, tentativeness, creativity, objectivity, subjectivity, social and cultural influences | High school, intro college | Silver | Elucidate students' views about several aspects of the nature of science. |
| Views on Science and Education (VOSE) | Nature of science, theories and laws, tentativeness, creativity, objectivity, | High school, intro college, intermediate, upper level | Silver | Create in-depth profiles of the views of students or adults about the nature of science and nature of science instruction. |

| | subjectivity, scientific method, teaching the nature of science | | | |
|---|---|---|---|---|
| **Self-Efficacy** | | | | |
| Sources of Self-Efficacy in Science Courses- Physics (SOSEC-P) | Self-efficacy | Intro college | Bronze | Assess students' beliefs that they can succeed in their physics course. |
| Physics Self-Efficacy Questionnaire (PSEQ) | Self-efficacy | Intro college | Bronze | Measure students' self-efficacy in their physics course. |
| Self-Efficacy in Physics Instrument (SEP) | Self-efficacy | Intro college | Bronze | Examine the relationship between physics self-efficacy and student performance in introductory physics classrooms. |

## A. Overview of Beliefs and Attitudes Assessments

There are 14 research-based assessments of students' beliefs and attitudes that we discuss here. There are four assessments about students' beliefs about learning physics in general: The Colorado Learning Attitudes about Science Survey[13] (CLASS), Maryland Physics Expectations Survey[14] (MPEX), Epistemological Beliefs Assessment for Physical Sciences[14,15] (EBAPS), and the Views About Science Survey[16,17] (VASS). There are five assessments about students' beliefs about specific aspects of physics or their own learning, e.g., labs, problem solving etc. These are the Colorado Learning about Science Survey for Experimental Physics[18] (E-CLASS), the Attitudes and Approaches to Problem Solving[19,20] (AAPS), Attitudes about Problem Solving Survey[21] (APSS), the Physics Goal Orientation Survey[22] (PGOS) and the Self Assessment of Learning Gains[23] (SALG). There are three assessments of students' views of their self-efficacy in their physics classes: Sources of Self-Efficacy in Science Courses- Physics[24] (SOSEC-P), Physics Self-Efficacy Questionnaire[25] (PSEQ) and the Self-efficacy in Physics Instrument[26] (SEP). There are two surveys of students' views about the nature of science. There are the Views on Science and Education Questionnaire[27] (VOSE) and the Views of Nature of Science Questionnaire[28] (VNOS). There are also additional assessments of motivation, discussed in Lovelace and Brickman[29] that may be of interest, but will not be discussed here.

Since these surveys of beliefs and attitudes don't assess the content covered in any course, they can be used at the high school level, and at all levels in the undergraduate and graduate curriculum (unless otherwise noted below). Many of these surveys can be used across disciplines or have versions specifically tailored to other disciplines. Most of these beliefs and attitudes surveys (unless otherwise noted) are meant to be given as a pre-test at the beginning of the semester and post-test at the end of the semester, in order to look at the shifts in beliefs scores during your course; they are also appropriate to give at other times in the semester (e.g., near exams) or across an entire course sequence.

### 1. Beliefs About Physics Learning in General

Many physics faculty care about their students learning to think like physicists, but often don't assess this because it is not clear how to do so best. Physics education researchers have created several surveys to assess one important aspect of thinking like a physicist: what students believe that learning physics is all about. These surveys are not about whether students like physics, but about how students perceive the discipline of physics or their physics course. These surveys measure students' self-reported beliefs about physics and their physics courses and how closely these beliefs about physics align with experts' beliefs.

The Colorado Learning About Science Survey[13] (CLASS - pronounced "sea-lass"), asks students to rank statements about their beliefs about physics and learning physics around such topics as real-world connections, personal interest, sense-making/effort and problem solving, using a 5-point Likert scale from strongly agree to strongly disagree. The survey is most commonly scored by to collapsing students' responses into a binary ("strongly agree" and "agree" are grouped, "strongly disagree" and "disagree" are grouped) depending on whether they are the same as an expert physicist would give and looking at the shift in student beliefs from pre-test to post-test. One would hope that after a physics course, students' beliefs would become more expert-like. The CLASS questions contain only one statement that students can agree or disagree with to help students interpret these questions consistently (as opposed to more than one idea in the same question). The CLASS questions were developed based on questions from the MPEX and VASS. The CLASS added questions about personal interest, aspects of problem solving and the coupled beliefs of sense-making and effort that were not included in the MPEX or VASS.[13]

The Maryland Physics Expectations Survey[14] (MPEX) measures students' self-reported beliefs about physics and their physics courses, their expectations about learning physics and how closely these beliefs about physics align with experts' beliefs. The surveys ask students to rank 5-point Likert scale questions about how they learn physics, how physics is related to their everyday lives, and about their physics course. Some of the MPEX questions are very course specific, e.g., they ask about a student's grade in the course. The format and scoring of the MPEX questions are the same as the CLASS questions. The questions on the MPEX were chosen through literature review, discussion with faculty, and the researchers' personal experiences.

The CLASS and MPEX are very similar and several items are the same on both tests. The MPEX and CLASS both ask questions about students' personal beliefs about learning physics, but the MPEX focuses more on students' expectations for what their specific physics course will be like. While the CLASS does not include questions about expectations for the specific course, it does include questions that only make sense in the context of a physics course, e.g., asking about students' belief that they can solve a physics problem after studying that physics topic. The MPEX takes longer to complete than the CLASS, presumably because the questions contain multiple ideas and are harder to parse. Both assessments have a strong research validation. The CLASS builds on the MPEX, and has been used more widely, so there is more comparison data available.[30]

The Epistemological Beliefs About Physics Survey (EBAPS) probes students' epistemology of physics, or their view of what it means to learn and understand physics.[14] The EBAPS also contains questions that are course specific (as opposed to being about learning physics in general), for example, one questions asks about how students should study in their physics class. The developers tried to ensure that the EBAPS questions don't have an obvious sanctioned answer and have a rich context in order to elicit students' views more successfully.[15] The EBAPS has three question types. Part one contains agree/disagree Likert scale questions, part 2 contains multiple-choice questions, and part 3 gives students two statements and asks them to indicate how much they agree with each (similar to the VASS). The level of sophistication of students' answers is scored using a non-linear scoring scheme where different responses have different weighting depending on how sophisticated the developers determined each answer was. The EBAPS is most appropriate for high school and college level introductory physics courses. The EBAPS questions were developed based on an extensive review of the MPEX and Schommer's Epistemological Questionnaire.[31] The developers synthesized other researchers' ideas to create guiding principles, which they used to write the EBAPS questions.

The main difference between the EBAPS and the CLASS and MPEX is the style of the questions, where the EBAPS has three styles of questions, and the MPEX and CLASS include only agree/disagree questions. The content on the EBAPS, MPEX, and CLASS is similar and all have high levels of research validation.

The Views About Science Survey[16,17] (VASS) is another survey for probing student beliefs about physics and learning physics. The VASS uses a special question format called contrasting alternative design where students compare and contrast between two viewpoints. For example, one question contains the statement "Learning in this course requires:" with the contrasting alternatives "(a) a special talent" and "(b) a serious effort." Students are asked to compare how much they agree with (a) and (b) by choosing between the following options: (a) >> (b), (a) > (b), (a) = (b), (a) < (b) or (a) << (b). Questions are scored in the same way as the MPEX and CLASS. The VASS can be used in introductory college physics courses as well as high school physics courses. VASS questions were developed based on an expert/folk taxonomy of student views about science.

The biggest difference between the VASS and the CLASS and MPEX is that the VASS uses the contrasting cases format. Because this format can be confusing to students if they don't agree that the answer choices given are opposites, and the expert-like response isn't always clear, the VASS is useful for discussing the ideas around students' beliefs about learning physics but is less useful for reliably measuring how expert-like your students' beliefs are. The CLASS and MPEX have more obvious expert-like answers, so their results can give you a better idea of how expert-like your students' views are. The content of the VASS is very similar to the CLASS and MPEX. Like the MPEX, the VASS has several questions that are course specific.

17. "Interpreting VASS Dimensions and Profiles for Physics Students," I. Halloun and D. Hestenes, Sci. Educ. **7**(6), 553–577 (1998). (E)

## 2. Beliefs About Physics Learning in a Specific Context

Beliefs/attitudes surveys have been created for three specific contexts: experimental physics (E-CLASS[18]), problem solving (AAPS[19] and APSS[21]) and goal orientations (PGOS[22]).

The Colorado Learning Attitudes about Science Survey for Experimental Physics[18] (E-CLASS) is designed to measure the influence of a laboratory course on students' epistemologies and expectations around experimental physics. The E-CLASS asks about a wide range of epistemological beliefs, so that it can be used in courses with a wide range of goals. The E-CLASS asks students to rate their agreement with statements by answering for themselves, "What do YOU think when doing experiments for class?" and answering for a physicist, "What would experimental physicists say about their research?" This helps instructors differentiate students' personal and professional epistemologies. The E-CLASS can be used in introductory, intermediate or upper-level laboratory courses. The E-CLASS score is calculated using the responses to the questions about students' personal beliefs (not the prompts about what they think a physicist's response is). The E-CLASS score is calculated by giving +1 point for an expert-like response (favorable), 0 points for a neutral response and -1 points for a novice-like response (unfavorable). The total score for the 30 questions can range from -30 to 30 points. You can also look at the shift in score from pre- to post-test to determine how the course influenced students' beliefs about experimental physics. The E-CLASS questions were developed based on consensus learning goals defined by faculty at the University of Colorado at Boulder for their laboratory curriculum. The questions were modeled after questions on the CLASS and based on common challenges instructors observed students having in laboratory courses.

There are two surveys that measure students' attitudes and approaches to problem solving in physics. These surveys are important because the way students think about problem solving can affect how they learn this skill, and faculty can target the development of problem-solving skills to help their students improve.

The Attitudes toward Problem Solving Survey[21] (APSS) is a survey of students' attitudes toward problem solving, e.g., how they think about equations, the process they go through to solve problems, their views on what problem solving in physics means, etc. Like other attitudes and beliefs surveys, students are asked to agree with statements using a 5-point Likert scale, strongly (dis)agree and (dis)agree are collapsed, and the percent expert response is calculated as the percentage of questions where students agree with the expert response. In addition to the agree/disagree questions, there are also two multiple-choice questions on the APSS. The APSS is appropriate for introductory college courses. Some of the APSS questions were adopted from the MPEX, while others were newly created.

Like the APSS, the Attitudes and Approaches to Problem Solving[19,20] (AAPS)

measures students' agreement with statements about their attitudes and approaches to problem-solving using a 5-point Likert scale. To calculate the average score for a question, +1 is assigned to each favorable response, -1 is assigned to each unfavorable response, and a 0 is assigned to neutral response, and the overall score is the average of the score for each question. The AAPS can be used at all levels of undergraduate courses, and at the graduate level.

Since the AAPS was developed by expanding the APSS, the topics covered and the questions on the AAPS and APSS are quite similar. Fourteen of the questions are the same or very similar between the tests. The AAPS has more questions (33 questions versus 20 questions), so it covers a few more aspects of problem solving than the APSS, including how students feel about problem solving, how they learn from the problem-solving process, use of pictures/diagrams, and what students do while solving a problem. The AAPS also includes questions that target graduate-level problem solving.

The CLASS, MPEX, EBAPS and VASS also contain questions about students' attitudes and beliefs about problem solving, similar to those on the APSS and AAPS. The AAPS and APSS can specifically target problem-solving beliefs, while the CLASS, MPEX, EBAPS and VASS ask about a wider range of beliefs and attitudes.

The Physics Goals Orientation Survey[22] (PGOS) is a survey of students' motivations and goal orientations in their physics course. These motivations can influence how students engage in their physics class and how well they learn the material. The PGOS addresses four goal orientations: task orientation (the belief that success is a product of effort, understanding, and collaboration), ego orientation (the belief that success relies on greater ability and attempting to outperform others), cooperation (when students value interaction with their peers in the learning process), and work avoidance (the goal of minimum effort – maximum gain). The PGOS uses a 5-point Likert scale, with 1 point given for strongly disagree, 5 points for strongly agree, and 2-4 points for disagree, neutral, or agree, respectively. The average score for each of the four goal orientations is calculated separately, and there is no overall score calculated. The PGOS is appropriate for introductory and intermediate university physics courses. It can be given as a pre- and post-test to determine how your course may have influenced students' goal orientations. The PGOS questions were taken from a previous survey of goal orientation by Duda and Nicholls[32] and revised so that they would be appropriate for a university-level physics course, with some new questions created. The PGOS was developed in Australia.

The Student Assessment of Learning Gains[23] (SALG) is an online assessment where students self-assess how different parts of their course impacted their learning using a 5-point Likert scale. It is like the student evaluation given at the end of most courses, but the questions only ask students about what they gained from different aspects of the course instead of what they liked. The SALG developers found that students' observations about what they gained from the class were useful to help faculty improve the course, whereas their observations about what they liked were not helpful.[23] You can use the SALG online system[33] to choose questions to include from each of the following categories: understanding of class content, increases in skills, class impact on attitudes, integration of learning, the class overall, class activities, assignments, graded activities and tests, class resources, the information you were given, and support for you as an individual learner. You can also edit and reorder questions. You can give the SALG at a midpoint in your class to get a sense of which parts of your course could be improved, or at the end to evaluate your students' understanding of how your course supported their learning. The SALG website[33] also has a "baseline instrument" available that can be used at the beginning of a course. The SALG was developed over 300 student interviews where students discussed what they had gained from certain aspects of a course, and what they liked.

### 3. Nature of Science

There are two main research-based surveys about the nature of science, the Views on Science and Education Questionnaire[27] (VOSE) and the Views about the Nature of Science Questionnaire[28] (VNOS), which probe students' views about the values and epistemological assumptions of science. These surveys can help faculty understand how their courses and teaching methods influence students' views of the nature of science. These can be especially useful in courses that aim to develop these views, such as courses for pre-service teachers. Both are intended as both a pre- and post-test.

The VOSE[27] is Likert-scale survey of students' beliefs about the nature science and beliefs about how you should teach the nature of science. The VOSE addresses seven major topics including tentativeness of scientific knowledge, nature of observation, scientific methods, hypotheses, laws, and theories, imagination, validation of scientific knowledge, objectivity and subjectivity in science. It also includes five questions about students' beliefs about teaching the nature of science. Each question consists of a question statement and 3-9 possible responses, with which students can agree or disagree with using a 5-point Likert scale. There are no right or wrong answers, but each statement corresponds to a particular "position" on one or more subtopics of nature of science. The developer has created an extensive list of coding categories to "create and in-depth profile of a [student's] nature of science views and educational ideas."[28] The coding categories can be found in Chen, 2006.[28] Burton[34] developed a numerical system for calculating a numerical score for each issue or topic, by assigning a number between 0 and 4 to a student's response for each item listed under that issue or topic and calculating the average. The VOSE can be used in high school courses and in introductory, intermediate, and upper-level undergraduate courses. The VOSE questions were developed based on questions from the Views on Science-Technology-Society[35] (VOSTS) and VNOS[28] to address concerns about the VOSTS and VNOS being open-ended and hard to administer and score. The VOSE aims to increase the validity of the survey and decrease interpretation biases.

The Views on the Nature of Science Questionnaire[28] (VNOS) is an open-ended survey of students' ideas about the nature of science, including the empirical, tentative, inferential, creative, theory-laden nature of science and the social and cultural influences on scientific knowledge. Many of the questions ask students to give an example to support their ideas. In addition to students written responses, the developers encourage faculty to do individual follow-up interviews with students to better understand the meanings of their responses to the questions. Students' responses can be scored as naïve, transitional, or informed based on a rubric for each question. The VNOS can be used with middle school, high school, and introductory college students. The VNOS questions were created by the developers and tested with students and experts.

The VNOS and VOSE cover similar topics around the nature of science. The main difference between them is the format. The VNOS is open-ended while the VOSE asks students to agree/disagree with different options. Because the VNOS is open-ended, it can be time consuming to score, and subject to interpretation bias, though conducting interviews with students about their responses reduces the chance of bias in scoring. Another difference between the VOSE and VNOS is that in addition to asking about students' philosophical beliefs about science, the VOSE asks students to agree/disagree with statements about how to teach the nature of science.

Many other multiple-choice instruments to assess students' views of the nature of science were developed in the 1960's, 70's, and 80's, but were based on researchers' ideas and not on student interviews or research into student thinking.[36] The VOSTS,[35] published in 1992, was the first nature of science instrument to use a student-centered design process, including analysis of student responses and student interviews. However, other researchers found many problems with students' interpretations of the VOSTS.[27,28,37] Both the VOSE and the VNOS were developed in response to these problems.

## *4. Self-Efficacy*

Self-efficacy is a person's situation-specific belief that they can succeed in a given domain.[38] There are three assessments that measure students' belief that they can succeed in their physics course. There are numerous other assessments of self-efficacy with differing focuses, e.g., other disciplines, self-efficacy in general, etc. We focus on those specifically developed for physics courses. All three of these assessments ask students to rate their agreement with statements on a five-point Likert scale and are appropriate for introductory college students.

The Sources of Self-Efficacy in Science Courses-Physics[24] (SOSEC-P) assesses students' beliefs that they can succeed in their physics course by asking them to agree or disagree with a series of statements. The questions are divided into four categories, corresponding to four established aspects of self-efficacy: performance accomplishment, social persuasion, vicarious learning, and emotional arousal. These questions ask about students' feelings about different aspects of the course, how the instructor and other students influenced their views of themselves, the students' behavior in the course (paying attention, working hard, etc.), and more. Several of the Likert-scale questions on the SOSEC-P were taken from existing math and general academic surveys of self-efficacy. Additional new questions were written based on the developers' experience with undergraduate science education.

The Physics Self-Efficacy Questionnaire[25] (PSEQ) is a similar survey of students' beliefs that they can succeed in their physics course. The PSEQ has five questions, so it probes only one dimension of self-efficacy. Specifically, the PSEQ focuses on students' confidence in their ability to succeed in their physics course. The questions do not mention specific portions of the course, or specific members of the course (other students, instructor, etc.). They simply ask the students about themselves and their own ability in their physics course. Most of the Likert-scale questions on the PSEQ are modified versions of questions from the General Self-Efficacy Scale,[39] while one PSEQ question was written by the developers. The PSEQ was developed in Australia.

The Self-Efficacy in Physics[26] (SEP) instrument is another survey that asks students to agree with statements about their beliefs about their ability to succeed in their physics course. The SEP contains 8 questions, which are more specific than those on the PSEQ. These questions ask students how good or bad they are at science/math, if they are good at using computers, and if they believe they can solve two specific mechanics problems. The SEP questions were developed based on a literature review and modeled after self-efficacy questions from surveys in other disciplines.

The SOSEC-P has 33 questions whereas the PSEQ and SEP have 5 and 8 questions respectively, so the SOSEC-P probes more dimensions of self-efficacy in more depth than the other surveys. There is a lot more variety in the questions on the SEP than the questions on the PSEQ. The SEP asks students about their belief that they can solve very specific physics problems, their comfort using a computer, and if they consider themselves good at math, whereas the PSEQ questions are about physics in general. All have the same level of research validation.

38.    "Self-efficacy: Toward a unifying theory of behavioral change,"  A. Bandura, in Advances in Behaviour Research and Therapy **1**(4), pp. 139–161 (1978). (E)

39.    Measurement of perceived self-efficacy: Psychometric scales for cross-cultural research,  R. Schwarzer, Freie Universität Berlin, Berlin (1993). (I)

**B.  Recommendations for Choosing a Beliefs and Attitudes Assessment**

*1. General Beliefs*

Use the CLASS if you want an assessment that is quick to complete, has a large amount of comparison data available and where the questions are easy for students to understand. Further, use the CLASS if you want to look at categories of questions that were determined through a rigorous statistical analysis, so they reflect students' views of the relationship between questions. The CLASS and MPEX statements refer to the kinds of activities that students do in a traditional introductory physics course, so the questions may not make sense to students if you are teaching in a very non-traditional way. If you have been using the MPEX, EBAPS or CLASS in the past, you may want to keep using these to compare your results. The MPEX was designed with a resources perspective, which assumes that students' ideas are not coherent, so if you want an assessment from the resources perspective, use the MPEX.

## 2. Specific Beliefs

Use the E-CLASS if you want to measure students' beliefs in the context of experimental physics. Use the APSS if you want to probe your students' attitudes about problem solving, including undergraduate and graduate students. Use the PGOS if you want to understand your students' motivations and goal orientations in their physics course. Use the SALG if you want to understand your students' perspective on which parts of your course helped them gain the most.

## 3. Self-efficacy

If you want to measure detailed changes in your students' physics course specific self-efficacy, use the SOSEC-P, as it probes several dimensions of self-efficacy and uses several questions to probe each. If you need a shorter self-efficacy assessment that can be combined with some other assessment, use the five-question PSEQ, which can give you a general sense of your students' belief and confidence in their ability in your course.

## 4. Nature of Science

Use the VOSE if you want a multiple-choice assessment that is quick and easy to score. Use the VNOS if you would like to use an open-ended survey to get a more detailed understanding of your students' views on the nature of science.

## V. PROBLEM SOLVING

**TABLE V.** Problem-solving assessments.

| Assessment | Focus | Intended population | Research validation | Purpose |
|---|---|---|---|---|
| Assessment of Textbook Problem Solving Ability (ATPSA) | Solving textbook problems | Intro college | Bronze | Gauge students' problem-solving ability in a first-semester calculus-based physics course. |
| Colorado Assessment of Problem Solving (CAPS) | Detailed understanding of students' problem solving | Graduate, upper-level, intermediate, intro college, high school, middle school | Bronze | Assess students' strengths and weaknesses on 44 different components of the problem-solving process, using a general problem-solving situation that is not tied to any specific discipline. |
| Minnesota Assessment of Problem Solving rubric (MAPS) | Rubric to score written problem solutions | High school, intro college | Silver | Assess written problem solutions on five different aspects of problem solving in undergraduate introductory physics courses. |

## 1. Overview of Problem-Solving Assessments

Students' ability to solve a problem when there is no solution method obvious to the solver[40] is a key skill many physics faculty would like their students to develop. Problem-solving can be defined in many ways, e.g., the ability to solve physics textbook problems,[41] or a collection of many components that a solver brings to bear to solve any problem, regardless of discipline.[42] Because of the variety of interpretations of what problem solving means, there are also a variety of instruments to measure different aspects of problem solving, including the Assessment of Textbook Problem Solving Ability[41] (ATPSA), the Colorado Assessment of Problem Solving[42] (CAPS) and the Minnesota Assessment of Problem Solving rubric[43] (MAPS). There are also surveys to probe students *attitudes* about problem-solving, rather than their skills (AAPS[19,20] and APSS[21]). These are discussed in the "Beliefs about Physics Learning in a Specific Context" above.

The Assessment Of Textbook Problem Solving Ability[41] (ATPSA) contains open-ended problems similar to end of chapter textbook problems. The content covered on the ATPSA is intentionally limited to Newton's Laws, energy, and momentum, as these are commonly taught topics in introductory courses. The ATPSA is meant for introductory undergraduate calculus-based mechanics courses and uses right/wrong grading, and can be given as a pre- and post-test, so the overall results can be used evaluate a course (but not individual students). The ATPSA can help instructors assess the impact of teaching reforms on students' ability to solve traditional physics problems. Basic algebra and trigonometry are required to solve the problems. There is a range of difficulty in the ATPSA questions so that the test can assess students of varying levels, though the level of math required for the questions does not change with the difficulty. There are no questions where a mathematical "trick" is needed. The questions on the ATPSA were created by the test developers.

The Minnesota Assessment of Problem Solving[43] (MAPS) rubric is a rubric that you can use to score your students' written solutions using 5 categories of problem-solving: 1) useful description, 2) physics approach, 3) specific application of physics, 4) mathematical procedures and 5) logical progression. The MAPS rubric is applicable to a wide variety of problem types and introductory physics topics. With this rubric, you score each student's written solution from 1 to 5 for each category, then look at the frequency of rubric scores for each category across the students in your class to get a sense of their problem-solving strengths and weaknesses. The MAPS rubric has been used at the high school and introductory college level. This rubric was created based on years of research on student problem solving at the University of Minnesota[44–46] and has been extensively studied for evidence for validity, reliability, and utility.[47]

The Colorado Assessment of Problem-Solving[42] (CAPS) is an open-ended problem-solving assessment which presents a general problem situation from the Jasper Woodbury Series[48] that is not tied to any specific discipline, so that students don't have to understand any particular physics concept in order to complete the assessment. The CAPS consists of a script describing a scenario and questions about how to solve the problems in that scenario. Students' responses to the questions are graded on a continuum using a rubric that assesses 44 different sub-skills of the problem-solving process, to gauge students' strengths and weaknesses in problem solving. There is no overall score, as the CAPS is meant to help you assess which aspects of problem solving an individual student needs more help with. It is appropriate for any level of student (middle school to graduate students). These 44 sub-skills are divided into three categories: (1) knowledge; (2) beliefs, expectations, and motivation; and (3) processes. Use it to give individual guidance to specific students, e.g., undergraduate research student, graduate student etc. It would not be appropriate to use to assess problem solving as a whole in your class.

## *2.   Recommendations for Choosing Problem-Solving Assessment*

If you want to assess your students' problem-solving skills on textbook-like problems that cover Newton's laws, momentum and energy, want something that is standardized so that you can compare over time and to others, and is reasonably easy to score, use the ATPSA. If you want to use a standardized method of scoring your students' written solutions to your own physics problems and want to get a better sense of your students' strengths and weaknesses with particular problem-solving skills, use the MAPS rubric. If you have a small number of students (undergraduate research students, graduate students, etc.), and you want to understand their problem-solving strengths and weaknesses in great depth, and you have time to individually go through the problem-solving exercise and associated questions with them, use the CAPS.

## VI. SCIENTIFIC REASONING

**TABLE VI.** Scientific reasoning assessments.

| Assessment | Focus | Intended population | Research validation | Purpose |
|---|---|---|---|---|
| Lawson Classroom Test of Scientific Reasoning (CTSR) | Proportional thinking, probabilistic thinking, correlational thinking, hypothetico-deductive reasoning | Intro college, high school, middle school | Gold | Measure concrete- and formal-operational reasoning. |
| Scientific Abilities Assessment Rubric (SAAR) | Represent information in multiple ways, design and conduct experiments, communicate scientific ideas, collect and analyze experimental data, evaluate experimental results | Intro college, high school | Silver | Assess students' scientific abilities as evidenced in their writing around experiments and design tasks. |

### 1. Overview of Scientific Reasoning Assessments

Scientific reasoning is an important skill that many faculty would like their students to develop. Most generally we can think of scientific reasoning skills as those needed to conduct scientific inquiry including evidence evaluation, inference and argumentation to form theories about the natural world.[49] There are two assessments of scientific reasoning that have been used in physics: the Lawson Classroom Test of Scientific Reasoning[50] (CTSR) and the Scientific Abilities Assessment Rubrics[51] (SAAR). The Physics Lab Inventory of Critical Thinking[52] (PLIC) also assesses aspects of students' scientific reasoning skills, but focuses more on their reasoning skills as related to labs and is discussed in the section titled, "Laboratory Skills."

The Lawson Classroom Test of Scientific Reasoning Ability[50] (CTSR) is a multiple-choice pre/post test with questions on conservation, proportional thinking, identification of variables, probabilistic thinking, and hypothetico-deductive reasoning. Lawson describes scientific reasoning as consisting of "a mental strategy, plan, or rule used to process information and devise conclusions that go beyond direct experience."[53] The CTSR was intended to help instructors classify students' reasoning abilities as concrete, transitional, or formal, based on the number of questions they answer correctly. Other instructors use the percentage correct on the CTSR for each cluster of questions to get a sense of their students' strengths and weaknesses around scientific reasoning. Those instructors don't classify each student as having a certain level of reasoning ability based on their total score for two reasons: it is not clear that the CTSR is measuring only one construct; and the validity and reliability of the most recent version of the CTSR have not been established. Most of the questions come in pairs, with one question about the answer and one about reasoning. The Lawson test was developed for high school students but has also been used at the introductory college level. The questions were originally based on demonstrations, where the instructor would perform the demonstration and then ask students questions about it in an interview format. The most recent version has converted these interview questions into a multiple-choice paper and pencil test.

The Scientific Abilities Assessment Rubrics[51] (SAAR) are a set of rubrics used to assess students' levels of competence around seven different scientific abilities:

1. The ability to represent physical processes in multiple ways

2. The ability to devise and test a qualitative explanation or quantitative relationship

3. The ability to modify a qualitative explanation or quantitative relationship

4. The ability to design an experimental investigation

5. The ability to collect and analyze data

6. The ability to evaluate experimental predictions and outcomes, conceptual claims, problem solutions, and models

7. The ability to communicate

The SAARs are used to assess specific scientific abilities as evidenced in students' written work around experiments or design tasks. The Scientific Abilities Assessment Rubrics outline the different levels of performance (0, missing; 1, inadequate; 2, needs some improvement; and 3, adequate) and include a description of each level, to enable students to self-assess as they work toward developing these abilities. In this way, the SAARs enable formative assessment of students' scientific abilities. Instructors can also use the SAARs to assess their students' acquisition of these scientific abilities by scoring students' laboratory write-ups for a particular experiment or design task from 0 to 3 using the descriptions of the different levels on the rubric. Instructors can then compare the distribution of scores for a particular scientific ability at the beginning and end of the course in hopes of seeing more students scoring "adequate." The SAARs were developed in the context of introductory college courses, though may also be appropriate for high school and intermediate college classes. The list of scientific abilities is based on literature on the history of the practice of physics, a taxonomy of cognitive skills, recommendations of science educators, and an analysis of science-process test items.

49. "The development of scientific thinking skills in elementary and middle school," C. Zimmerman, Dev. Rev. **27**, 172–223 (2007). (E)

50. "The development and validation of a classroom test of formal reasoning," A. E.

### *Recommendations for Choosing Scientific Reasoning Assessment*

Use the CTSR if you want to assess your students' reasoning level, possibly in conjunction with an appropriate test of their mathematical skill or physics content knowledge. Because it reduces students' reasoning to one of three levels, which are assumed independent of content, it presents a rough guideline for how your class is doing as a whole. Do not use this test if you need more detailed information about a specific student or group of students (such as for placement into a particular class), because the assumption that reasoning level is content-independent can break down when you look at individuals.

Use the SAARs to help your students self-assess their scientific abilities in lab courses. You can also use the SAAR as an instructor to give your students feedback on their competency around specific scientific abilities and sub-abilities and look at how your students' scores change over the course of your class.

## VII. LABORATORY SKILLS

**TABLE VII.** Laboratory skills assessments.

| Assessment | Focus | Intended population | Research validation | Purpose |
|---|---|---|---|---|
| Physics Lab Inventory of Critical Thinking (PLIC) | Evaluating experimental methods, generating and evaluating conclusions based on data, comparing measurements with uncertainty, evaluating data fitted to a model, | Upper-level, intermediate, intro college | Silver | To assess how students critically evaluate experimental methods, data, and models. |
| Concise Data Processing Assessment (CDPA) | Measurement and uncertainty, relationships between functions, numbers and graphs. | Graduate, upper-level, intermediate, intro College | Silver | To probe student abilities related to the nature of measurement and uncertainty and to handling data. |
| Physics Measurement Questionnaire (PMQ) | Measurement and uncertainty | Intro college | Silver | To probe students understanding of measurement and uncertainty using open-ended sample discussions. |
| Measurement Uncertainty Quiz (MUQ) | Measurement and uncertainty | Intro College | Bronze | To discuss measurement and uncertainty concepts with students, and why one answer might be better than the others. |

*1. Overview Laboratory Skills Assessments*

Faculty often assume that during the laboratory portion of a physics course, students develop the ability to gather and evaluate data through experiments. Several assessments of different aspects of lab skills have been created to help instructors evaluate their students' laboratory skills and critical thinking ability at the beginning and end of the course. There are four assessments of lab skills, the Physics Lab Inventory of Critical Thinking[52] (PLIC), the Concise Data Processing Assessment[54] (CDPA), the Physics Measurement Questionnaire[55,56] (PMQ) and the Measurement Uncertainty Quiz[57] (MUQ). There is also an assessment to gauge students' attitudes about experimental physics (E-CLASS[18]), which is discussed in the section titled, "Beliefs About Physics Learning in a Specific Context." The Data Handling Diagnostic[58] (DHD) is another assessment of laboratory skills, which will not be discussed further here because the authors did not finish the development and validation of this assessment, and advise others to use the CDPA instead of the DHD.

The Physics Lab Inventory of Critical Thinking[52] (PLIC) assesses the way students critically evaluate experimental methods, data, and models and is the newest laboratory skills assessment. The PLIC includes an introduction that describes an experiment using masses and spring, and sample lab notebook entries for two groups of physicists. The PLIC uses "choose many" multiple-choice questions as well as Likert-scale questions to assess students' critical thinking around the lab notebook entries for this experiment. Students' responses are compared to the "expert-like response" for each question. Because many of the multiple-choice questions allow students to "choose many," the score for each question is calculated as the fraction of students who choose *at* *least one* of the expert-like responses. The PLIC has been used in all levels of undergraduate labs. The PLIC is still under development. The questions on the PLIC were based on the series of questions an expert posed to himself when conducting an introductory physics experiment.

The Concise Data Processing Assessment[54] (CDPA) is a 10 question multiple-choice pre-post assessment that measures students' understanding of handling data with questions around uncertainty in measurements and the relationships between functions, graphs and numbers. The CDPA is appropriate to use in any lab course with learning goals around data handling. The questions were based on established learning goals for an introductory laboratory course and iteratively refined using student interviews, expert review, and statistical analyses.

Both the PLIC and the CDPA assess students' understanding of data analysis skills, but the PLIC also assesses other skills including how students critically evaluate experimental methods, data, and models. The CDPA has 10 multiple-choice questions, where each has its own context, whereas the PLIC has one rich experimental context outlined at the beginning of the assessment, to which all 16 questions refer. Both the PLIC and CDPA have strong research validation.

The Physics Measurement Questionnaire[55,56] (PMQ) is an open-ended pre/post assessment of students' understanding of experimental measurements, including data collection, data processing and data set comparison. There is an experimental situation described at the beginning of the assessment and all the questions refer to this same experimental situation (similar to the PLIC). The questions ask students to reflect on how many measurements they should take, how to report the results of multiple measurements, how to compare sets of measurements, and how to fit a line to experimental data. Because the PMQ questions are open-ended, the answers and explanations are coded according to an established coding scheme, which can be time consuming. In each question or "probe," there is a short conversation between several people, and students are asked to choose which they most agree with, and then give a written explanation for their choice. The discussions in each probe are written with concise, simple language in order to be understandable for a wide range of English language levels. The developers use the PMQ results to look at their students' paradigms of measurement as either, "point" or "set." A point paradigm would see each measurement as the possible true value, where differences between measurements are a result of environmental factors or experimenter mistakes.[59] In the "set" paradigm, each measurement is an approximation of the true value, and deviations are random and always present. A set of measurements yields the best approximation of the true value, with an associated uncertainty. The questions on the PMQ were based similar questions from the Procedural and Conceptual Knowledge in Science (PACKS) Project.[60]

The PMQ has a unique format compared to the other laboratory skills assessments, where each question includes a conversation between students, with an open-ended question about the conversation. Further, the scoring of the PMQ is different from the other assessments discussed, and instructors code the responses to understand their students' results. The content and skills assessed on the PMQ are also included in the PLIC, though the PLIC goes into more depth in asking students to evaluate critically experimental methods, data, and models.

The Measurement Uncertainty Quiz[57] (MUQ) is a non-standard assessment that can be used as the basis of a discussion about precision, significant figures, accuracy, and error propagation with your introductory physics students. The developer explains that it is difficult to create a right/wrong test around the topics of measurement and uncertainty, because even experts may disagree on the correct answer. Because of this limitation, the MUQ questions are an opportunity to discuss with your students why one answer may be better than others. Because the MUQ is for discussion (and is not scored), it is not given as a pre/post-test. The 10 questions on the MUQ are a sample of the open-ended questions given to approximately 100 introductory physics students and 30 experts (graduate physics students and teachers). The most common responses were edited and turned into the multiple-choice options.

The MUQ focuses just on measurement uncertainty, whereas the CDPA also asks about fitting data, and relating functions, graphs and numbers. Both tests use the same question format and have the same number of questions, but the MUQ developers recommend using it to have a discussion with students, instead of using it as a pre/post test and scoring it, as you would with the CDPA.

54.    "Development of the Concise Data Processing Assessment," J. Day and D. Bonn, Phys. Rev. Spec. Top. - Phys. Educ. Res. **7**(1), 010114 (2011). (E)

55.    "Point and set reasoning in practical science measurement by entering university

### *2. Recommendations for Choosing Laboratory Skills Assessment*

Use the PLIC to get a rich understanding of your students' skills around critically evaluating experimental methods, data, and models. The PLIC assesses the content covered on the MUQ, CDPA, and PMQ, and additional content and skills related to critical thinking around experimentation. If you want a short, simple multiple-choice test of measurement uncertainty and relationships between functions, data, and graphs, use the CDPA. If you are interested in understanding your students' open-ended responses about data collection, processing and comparison, or in looking at your students' paradigms of measurement as either, "point" or "set," use the PMQ. If you want to have a rich conversation about measurement uncertainty with your students, use the MUQ as the basis of the conversation.

## VII. Observation Protocols

**TABLE VIII.** Observation protocols

| Title | Focus | Intended pop. | Research validation | Purpose |
|---|---|---|---|---|
| Teaching Dimensions Observation Protocol (TDOP) | Instructor and student classroom behaviors | All levels | Gold | To classify instructor and student behaviors in STEM classrooms. |
| Reformed Teaching Observation Protocol (RTOP) | Degree of reformed teaching | All levels | Gold | To assess the degree to which reformed teaching is occurring in classrooms. |
| Classroom Observation Protocol Undergraduate STEM (COPUS) | Instructor and student classroom behaviors | All levels | Silver | To classify instructor and student behaviors in STEM classrooms. |
| Real-time Instructor Observation Protocol (RIOT) | Instructor-student classroom interactions | All levels | Silver | To classify instructor interactions with students in STEM classrooms. |
| Behavioral Engagement Related to Instruction (BERI) | Student engagement | All levels | Bronze | To quantitatively measure student engagement in large university classes |
| Laboratory Observation Protocol for Undergraduate STEM (LOPUS) | Instructor and students' lab behavior | All levels | Bronze | To classify instructor and student behaviors in STEM labs. |
| Student Participation Observation Tool (SPOT) | Instructor and student classroom behaviors | All levels | Bronze | To classify instructor and student behaviors in STEM classrooms. |

## 1. Overview of Observation Protocols

Faculty in physics departments often observe each other's teaching and give each other feedback to improve teaching. Using an observation protocol for these informal observations can help faculty articulate the goals of these observations and focus on particular aspects of the classroom. Observation protocols can provide data that illustrate what happened in the class, which can be useful for self-reflection and professional development. You can use observation protocols once as stand-alone activity, or to track your own improvement.

Observational protocols can be conducted using segmented, continuous, and holistic procedures.[61] Segmented protocols are those in which the class period is broken up into shorter periods of time, two-minute intervals for example, and then observers note whether they see certain behaviors during that interval or not. At the end of the observation, observers note the number of intervals in which each of the different behaviors happened. Continuous protocols allow observers to indicate what is happening at any given moment in a class, and an observation results in a time-line indicating what happened when. This also allows the different classroom activities to be considered as a certain percentage of overall class-time. Holistic protocols are protocols in which the entire course is considered at once. This is done using a series of questions that the observer responds to at the end of an observation.

There are seven observation protocols that we will discuss here. Four of these protocols focus on recording what is happening in the classroom. These are the Classroom Observation Protocol Undergraduate STEM[62] (COPUS), the Teaching Dimensions Observation Protocol[63,64] (TDOP), the Real-time Instructor Observation Protocol[65,66] (RIOT) and the Student Participation Observation Tool[67] (SPOT). One protocol, the Laboratory Observation Protocol for Undergraduate STEM[61] (LOPUS), focuses on recording what is happening in laboratory courses. One protocol, the Reformed Teaching Observation Protocol[68] (RTOP) focuses specifically on assessing the degree of reformed teaching. Finally, one protocol, the Behavioral Engagement Related to Instruction[69] (BERI) looks at the level of student engagement in a class session. All of these observation protocols can be used in high school or college-level courses.

Perhaps the most well-known observation protocol is the Reformed Teaching Observation Protocol[68] (RTOP), a holistic paper and pencil observation protocol developed to evaluate the extent to which a classroom uses reform-based teaching techniques. The RTOP consists of 25 Likert-scale items from three different categories including, "lesson design and technique," "content," and "classroom culture." Observers watch a class session and respond to each item with a maximum of 4 meaning the item is "very descriptive" of the class, to a minimum of zero indicating that item "never occurred." The RTOP data can be reduced to a single score by adding up the scores for each item. A higher RTOP score means that a class is more reformed. The single RTOP score makes it particularly useful as quantitative evidence of instructor change in practice over time. There are several questions on the RTOP that evaluate the instructor on the content or design of the lesson, so it is more appropriate to use the RTOP with instructors that designed the lesson themselves (and not a teaching assistant who did not have autonomy in deciding what happens in the classroom). The RTOP developers emphasize that RTOP results are not valid unless the observers have gone through several days of training on how to use the instrument. The items on the RTOP were developed based on previous research and existing instruments.[68]

The Teaching Dimensions Observation Protocol[63,64] is a segmented observation protocol that aims to record what is happening in the classroom, unlike the RTOP, which is designed to evaluate the degree of reformed teaching. The TDOP looks at three basic dimensions of the classroom including "instructional practices," "student-teacher dialogue," and "instructional technology," and three optional dimensions, including, "potential student cognitive engagement," "pedagogical strategies," and "students' time on task." Each of these dimensions have codes associated with them, and observers memorize the meaning of these codes (28 basic, and 11 optional), and circle that code when it happens during each two-minute interval of an observation. Observers can collect data with pencil and paper, or with a computerized interface available on the TDOP website.[70] Once data are collected, observers can examine the percentage of intervals that each code (or code category) appears. The TDOP website also automatically creates some charts and graphs for review. TDOP creators recommend that users establish inter-rater reliability, and stress that training may take several days depending on how many dimensions are used. Both a TDOP users guide and TDOP scoring sheet are available for download.[70] The codes and categories on the TDOP were developed based on an instrument designed to study inquiry-based middle school sciences courses.[71]

The Classroom Observation Protocol for Undergraduate STEM[62] (COPUS) was developed based on iterative modifications of an early version of the TDOP, so it is also a segmented protocol and is similar in many ways. The COPUS developers aimed to create a more user-friendly version of the TDOP (though the version of the TDOP they were working with had more mandatory categories, and the newer version of the TDOP discussed in this paper has been simplified). COPUS codes are separated into two broad categories, "what teachers are doing," and "what students are doing," with a total of 25 codes using simplified language. This allows individuals to learn to use COPUS much more quickly than the TDOP or RTOP, in as few as 1.5 hours. The COPUS developers also added some categories that were aligned with best practices in large-enrollment college-level STEM, such as discussions motivated by clicker questions. Like the TDOP, observers indicate whether a certain behavior happened or not in each 2-minute period using a specialized scoring sheet. The COPUS developers have also recently developed the COPUS profiles online tool,[72] that allows a user to upload COPUS data in a spreadsheet in order to create several different visual representations of these data that can be helpful for reflection.

The Real-time Instructor Observation Protocol[65,66] (RIOT) was developed independently from COPUS at the same time, and therefore the two were developed to fill similar needs but with slightly different focuses. RIOT, which is similar to COPUS and TDOP, allows an observer to categorize what is happening during a classroom observation. Unlike the COPUS and TDOP, the RIOT is a continuous protocol that only follows the instructor and records what they are doing (including if they are interacting with students) but does not record what students are doing independently of the instructor. The categories for RIOT are organized by the types of interactions that are possible with students in the classroom, "talking at students," "talking with students," "observing students," and "not interacting with students." The RIOT was originally developed as a part of a Teaching Assistant (TA) pedagogy course to help new graduate student TAs understand how to interact with students in an active learning environment, so it is useful for helping faculty as well as teaching assistants understand and improve their teaching. Like COPUS, RIOT, requires little training to use. The RIOT categories were developed based on observations of classrooms using the CLASP curriculum[73] at University of California at Davis, and emergent behaviors seen there.

The Student Participation Observation Protocol[67] (SPOT) is an observation protocol very similar to the RIOT in that it is web-based and continuous and has the same developers, but there are a few key differences in the content and layout. SPOT had a more rigorous development process than RIOT, as categories are backed by research on student-centered learning in science classrooms. SPOT categories represent the observable features of seventeen of the best practices in active learning.[67] Different from RIOT, the SPOT records what both the instructor and students are doing (whereas RIOT focuses on the instructor), and is organized by class "mode" referring to how the instructor and students are interacting with each other at any given time. The class can be in "small group mode," where students are working in small groups, "whole class mode," where students are watching a lecture, movie, or demo, and "independent mode" where students are working silently and independently (such as when they are taking an exam). In each mode, different codes are available to describe different behaviors of instructors and students. SPOT is optimized for courses that include some traditional lecture elements in order to better classify how participation happens, and who is participating. For example, during a lecture where the instructor may interact by asking or answering questions, SPOT allows an observer to classify student responses as either shouted-out, asking a question, answering a question, contributing an idea, or via whole-class choral response. SPOT also allows the observer to keep track of individual students using a map interface based on where they are sitting in the room. This can help instructors determine if many students are participating, or if it is the same five or six each time. Since SPOT is web-based like RIOT, it also generates colorful figures useful for self-reflection.

The Laboratory Observation Protocol for Undergraduate STEM[61] (LOPUS) was developed to categorize student and teacher actions in laboratory settings. The LOPUS creators started their development with a draft of the COPUS, then reviewed the literature and watched video of laboratory classes, to determine new behaviors that should be added to the LOPUS, that were not included in the COPUS. Like the COPUS, the LOPUS is a segmented protocol and organized into two broad categories of instructor behaviors and student behaviors, but LOPUS also has a third category that captures the content of student and teacher verbal interactions in laboratory classes, and who (teacher or student) initiated the interaction. For example, someone viewing an instructor lecturing about data analysis would use the pair of codes: "Lec" (indicating that the instructor is lecturing) and a qualifying code from this third category, "Ana" (indicating that the conversation is about data analysis and calculations). The LOPUS team also cut some of the codes from the COPUS that they found were highly correlated to each other, in order to cut back on the number of codes an observer needed to memorize. The LOPUS is available in a web-based format through the General Observation and Reflection Platform (GORP).[74] The platform auto-creates charts and plots that are useful for reflection.

The Behavioral Engagement Related to Instruction[69] (BERI) protocol is a segmented observation protocol to measure student behavioral engagement, defined as on-task behavior, in large university classes. The BERI can help an instructor figure out which parts of their class resulted in higher student engagement. The BERI protocol outlines six engaged behaviors, for example, listening, writing, and engaged instructor interactions, and six unengaged behaviors, for example, settling in/packing up, being off-task or disengaged computer use. The observer chooses a group of ten students and sits near them. During the class, the observer cycles through each of the 10 students and records, on a printout of instructor notes, if each student was engaged or disengaged during part of the class. The BERI observation protocol categories were developed based on observations of large classes to determine which student behaviors could be defined as engaged and disengaged.

The BERI protocol focuses particularly on student engagement, whereas the other protocols discussed above have a more general focus. The COPUS, LOPUS, TDOP, and SPOT all record student behaviors during the class, but they do not label these behaviors as engaged or disengaged. Use the BERI if you are particularly interested in how your students' level of engagement in class depends on what you are doing in class.

Several of the observation protocols mentioned here have been incorporated into web-based tools to make them easier to use. See our article on PhysPort[75] for more information on accessing these online protocol tools.

### *Recommendations for Choosing an Observation Protocol*

While all the observation protocols discussed here are potentially useful for self-reflection and professional development, we particularly recommend the COPUS and RIOT for these purposes based on their short training times, and resources for self-training. Use the COPUS for professional development if you are interested in what specific pedagogical tools used in the classroom (ex: students making a prediction, instructor showing a demos). Use RIOT if you are more concerned with what the instructor is doing, and their interactions (ex: instructor is explaining content, instructor is listening to a question). If you are interested in evaluating how reformed a course is, especially if you want to apply a numeric score to this evaluation, and are able to attend a training, use the RTOP. If you want a detailed account of what pedagogical actions take place in a classroom, and have time for training, use the TDOP. Use SPOT if you have questions about the frequency, type, and diversity of student participation in the classroom. Use the LOPUS if you are interested in lab environments and use the BERI if you are particularly interested in how your students' level of engagement in class depends on what you are doing in class.

## VIII. Survey of Faculty Teaching Practice

There are two surveys of faculty instructional practices that are commonly used in physics, The Teaching Practices Inventory[76] (TPI) and the Postsecondary Instructional Practices Survey[77] (PIPS). Both are research-based surveys that ask faculty to self-report on their teaching and the kinds of things that go on in their classrooms and the durations. Researchers use these surveys to characterize the self-reported teaching practices of faculty, though the results could also be used by faculty themselves to illustrate their teaching practices in tenure and promotion documents. Since this Resource Letter focuses on assessments that faculty can use to understand and improve their own classroom, we will not discuss surveys that faculty can use to report on their own teaching practice in more detail.

## IX. CONCLUSION

This paper summarizes major RBAIs in non-physics-content areas: mathematics (Table III), beliefs and attitudes (Table IV), problem solving (Table V), scientific reasoning (Table VI), lab skills (Table VII), and observation protocols (Table VIII). In contrast with the previous Resource Letter in this series (RL:RBAI-1), this collection of RBAIs is generally used to augment our picture of student learning in physics rather than investigate their understanding of specific physics topics. RBAIs in this collection are useful at all points in the high school and undergraduate curriculum.

**ACKNOWLEDGMENTS**

We gratefully acknowledge the contributions of the other members of the PhysPort team and KSUPER who worked on this project: John D. Thompson, Jaime Richards, Devon McCarthy, and Brian Danielak. This work was partially supported by NSF grants PHYS-1461251, DUE-1256354, DUE-1256354, DUE-1347821, and DUE-1347728.